# MODELLING STRUCTURAL BREAKS IN STOCK PRICE TIME SERIES USING STOCHASTIC DIFFERENTIAL EQUATIONS

*D. Karzanov*

*National Research University Higher School of Economics, Moscow*

This paper studies the effect of quarterly earnings reports on the stock price. The profitability of the stock is modelled by geometric Brownian diffusion and the Constant Elasticity of Variance model. We fit several variations of stochastic differential equations to the pre-and after-report period using the Maximum Likelihood Estimation and Grid Search of parameters method. By examining the change in the model parameters after reports' publication, the study reveals that the reports have enough evidence to be a structural breakpoint, meaning that all the forecast models exploited are not applicable for forecasting and should be refitted shortly.

*List of keywords:* stock market, earnings reports, financial time series, structural breaks, stochastic differential equations.

## 1. Introduction

The paper aims to consider the publication of quarterly earnings reports as a point of structural change in the stock price time series. A quarterly earnings report is a filing released by public firms to report their performance. They consist of a balance sheet, cash-flow statement, and income statement. By studying reports, investors can assess the financial health of the firm and decide if it deserves their investment or not.

We discuss a way of describing the price time series not by a parametric distribution but by a stochastic differential equation (SDE). It will allow us to make a mathematical formulation through a variational statement to reveal when the parameters producing the series move from one principal component to another principal component. So, when the nature of the series changes (this is not a smooth change), it is called a structural break.

If we succeeded in writing the stock price (or profit) in the form of a differential equation, then we would add more value to the effort that many econometricians had made in the works dedicated to structural break identification using statistical tests. SDEs would move us towards a more correct, mathematically formulated problem. Many have done various empirical studies and now we should add a mathematical description of the problem.

## 2. Data

First things first, we need to say a few words about the data that we will study, as well as mention its sources. We collect the information about the companies that reported on the given date using the *yahoo_earnings_calendar* module in *Python*. The output contains the most important information after the report in tabular form. The companies are obliged to report while the market is closed. It is important to understand that a firm may provide the report before market opening or after market closure (e.g. before 09:30 or after 16:00 for NASDAQ). Therefore, if a firm publishes a report after market closure, we add one day to the report day.

We do not wish to confuse a report's effect on a ticker with an effect after weekends when the stock market had two days off. Many tickers have a tendency to soar or plunge on Mondays and then stabilize. Therefore, we exclude the tickers that were reported on Monday. Also, we want the market to be open after the report for some continuous period (for at least two working days), so we do not consider Friday's reports as well.

As soon as we collect the tickers that reported in April 2021, we are ready to move to the price collection. To download the prices from the target period we use different APIs such as *yfinance* from yahoo finance and *IEX*. We request the prices for one week before and after the report with a 5-minute interval.

# 3. Modelling

## 3.1 Geometric Brownian Diffusion

To begin with, we add a mathematical description of the problem. Usually, financial analysts and quant-researchers apply stochastic differential equations to describe the profitability of a stock. Our base is a diffusion process, namely geometric Brownian motion from Black&Scholes model:

$$\frac{dS_t}{S_t} = \mu dt + \sigma dW_t$$

where $S_t$ is the stock price, $\mu dt$ is the trend component, $\sigma dW_t$ is the risk component.

If we manage to estimate these two parameters using the data before and after the report release, we can compare quantitatively the change caused by a structural break.

Since the equation has a solution, we can use Maximum Likelihood Estimators (MLE) to find the estimates. We fit the model to a week before and after the report (left and right further on) data, and compare the values. A similar problem was discussed by Jakob Croghan (2017), where the GBM estimators for oil price were calculated using the following formulas:

$$\bar{X} = \frac{1}{n}\sum_{t=1}^{n} \log\left(\frac{X_t}{X_{t-1}}\right) \qquad \hat{\mu} = \bar{X} + \frac{\hat{\sigma}^2}{2}$$

$$\hat{\sigma} = \sqrt{\frac{1}{n-1}\sum_{t=1}^{n}\left(\log\left(\frac{X_t}{X_{t-1}}\right) - \bar{X}\right)^2}$$

given that $X_t$ is oil price at time $t$. After fitting this model to two subsets and comparing the estimates, we obtain table 1. An example of a fitted curve is shown in figure 1.

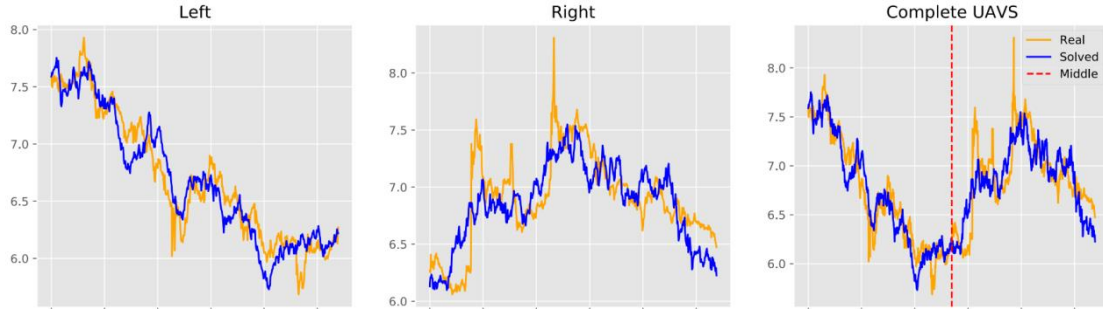

**Fig. 1.** MLE SDE fitted to UAVS prices. UAVS - AgEagle Aerial Systems Stock, report date: 2021-03-21. A red dashed line - report publication

**Table 2.** Numeric results from MLE for 20 tickers. The figures show the percentage change and the right-to-left ratio of the parameters for μ and σ

| ticker | mu change % | mu ratio r/l | sigma change % | sigma ratio r/l | ticker | mu change % | mu ratio r/l | sigma change % | sigma ratio r/l |
|---|---|---|---|---|---|---|---|---|---|
| AIKI | -339.96 | -2.40 | -36.32 | +0.64 | WAFU | -121.22 | -0.21 | -75.55 | +0.24 |
| INPX | +16.72 | +1.17 | -24.58 | +0.75 | GBOX | +16.82 | +1.17 | +34.85 | +1.35 |
| AACG | +18,579.93 | +186.80 | -36.46 | +0.64 | HITIF | -993.72 | -8.94 | +7.49 | +1.07 |
| SILV | -170.23 | -0.70 | +4.17 | +1.04 | PRPO | +538.48 | +6.38 | -9.81 | +0.90 |
| GTEC | +414.21 | +5.14 | +0.81 | +1.01 | NEXI | -1,209,087.14 | -12,089.87 | +2.27 | +1.02 |
| TRTC | +134.93 | +2.35 | +5.91 | +1.06 | VISL | -22.82 | +0.77 | -24.51 | +0.75 |
| SPRT | -91.60 | +0.08 | -86.49 | +0.14 | RXMD | -168.20 | -0.68 | +19.82 | +1.20 |
| MMEDF | -205.15 | -1.05 | -6.50 | +0.93 | CWCO | +25.87 | +1.26 | -9.90 | +0.90 |
| UAVS | -140.74 | -0.41 | +28.90 | +1.29 | MLSS | +30.38 | +1.30 | -10.01 | +0.90 |
| ESGC | -64.26 | +0.36 | -34.42 | +0.66 | PLNHF | -90.02 | +0.10 | -22.38 | +0.78 |

Table 3. Numeric results using CEV model with jump-diffusion modification

| ticker | mu change % | mu ratio r/l | sigma change % | sigma ratio r/l | jump % | gamma change % | gamma ratio r/l |
|---|---|---|---|---|---|---|---|
| AIKI | +221.41 | +3.21 | -88.55 | +0.11 | +0.00 | +36.36 | +1.36 |
| INPX | -37.90 | +0.62 | -24.58 | +0.75 | +0.00 | +57.14 | +1.57 |
| AACG | +30.41 | +1.30 | -94.24 | +0.06 | +0.00 | +400.00 | +5.00 |
| SILV | -187.40 | -0.87 | -83.32 | +0.17 | +0.00 | +1,200.00 | +13.00 |
| GTEC | +160.87 | +2.61 | -60.77 | +0.39 | +10.00 | +133.33 | +2.33 |
| TRTC | +173.90 | +2.74 | +1,068.47 | +11.68 | +10.00 | +1,900.00 | +20.00 |
| SPRT | -79.86 | +0.20 | -95.40 | +0.05 | -10.00 | +266.67 | +3.67 |
| MMEDF | -163.17 | -0.63 | +419.90 | +5.20 | +10.00 | -65.00 | +0.35 |
| UAVS | -134.12 | -0.34 | -44.25 | +0.56 | +0.00 | +0.00 | +1.00 |
| ESGC | -60.83 | +0.39 | -77.79 | +0.22 | +0.00 | +400.00 | +5.00 |
| WAFU | -139.48 | -0.39 | -59.76 | +0.40 | +10.00 | -18.18 | +0.82 |
| GBOX | -6.04 | +0.94 | +101.81 | +2.02 | -10.00 | +0.00 | +1.00 |
| HITIF | -25,119.13 | -250.19 | -67.03 | +0.33 | +0.00 | -76.67 | +0.23 |
| PRPO | +377.22 | +4.77 | -31.82 | +0.68 | +0.00 | +28.57 | +1.29 |
| NEXI | -178.04 | -0.78 | +27.24 | +1.27 | -10.00 | +57.14 | +1.57 |
| VISL | -44.08 | +0.56 | +113.78 | +2.14 | -10.00 | -85.71 | +0.14 |
| RXMD | -412.55 | -3.13 | +819.99 | +9.20 | +0.00 | +53.85 | +1.54 |
| CWCO | -11.48 | +0.89 | -77.21 | +0.23 | +0.00 | +600.00 | +7.00 |
| MLSS | +52.73 | +1.53 | +72.07 | +1.72 | +0.00 | -35.00 | +0.65 |
| PLNHF | -111.89 | -0.12 | +178.17 | +2.78 | -10.00 | -85.71 | +0.14 |
| RCON | -248.90 | -1.49 | -40.40 | +0.60 | +0.00 | +40.00 | +1.40 |

## 3.2 Constant Elasticity of Variance

Another model which can be applied in our settings is CEV formulated by Cox, J (1975) as

$$dS_t = \mu S_t dt + \sigma S_t^\gamma dW_t$$

Parameter $\gamma$ controls the volatility. $\gamma > 0$ means that in the given markets volatility rises when the price of the stock rises. Unfortunately, the equation does not have a closed-form solution, so we need to use another way of estimating its parameters. A computationally expensive but effective method is Grid Search which considers every combination of parameters and finds which set results in the lowest error. As a minimization criterion, we tried different metrics such as Mean Absolute Percentage Error and Kullback–Leibler divergence (relative entropy). However, the most reasonable fit was demonstrated by Mean Squared Error, which we consequently applied to all models in the summary.

## 3.3 Grid Search

The main problem with this approach is computational expenses - even for 20 drift values and 30 volatility values, the algorithm has to consider 600 options for one series. Another issue that we encounter is the determination of the grid range for parameters knowing the array of points only. If a search interval does not include the true parameter value, the results will likely be inferior. We propose using the following intervals for trend and volatility. Given that $y_i$ is the stock price at time $i$ and $z_i = y_i - y_{i-1}$:

$$\hat{\mu} = \frac{1}{n}\sum_{i=2}^{n}(y_i - y_{i-1}) \qquad [\hat{\mu} - 5 \cdot s.d.(z/n), \hat{\mu} + 5 \cdot s.d.(z/n)]$$

We faced many difficulties while choosing the correct borders for the volatility interval. It appeared to be a more complicated task because the interval is not necessarily symmetric around the point estimate. After all, the volatility cannot be negative, so the interval is $[\hat{\sigma}/10, \hat{\sigma} \cdot 25]$. As a point estimate for $\sigma$, we use the estimation of oil price volatility described by Croghan J. (2017):

$$\hat{\sigma} = \sqrt{\frac{1}{n}\sum_{t=1}^{n}\left(\log\frac{y_t}{y_{t-1}} - \bar{y}\right)^2} \qquad \bar{y} = \frac{1}{n}\sum_{t=1}^{n}\log\frac{y_t}{y_{t-1}}$$

### 3.4 Jump Diffusion

After analysing the indices, we noticed that the series demonstrated a leap or a jump which is not considered in the initial model. The quants usually add a jump component to the theoretical equation, which is modelled using Poisonous-like distributions:

$$\frac{dS_t}{S_t} = \mu_t dt + \sigma dW_t + (Y_t - 1)dN_t$$

In our implementation, we use the discussed CEV model, but consider three different initial values for the equation. For the right model, the last point of the left subsample is taken as the initial value scaled by jump parameter, $Y_t, \in \{.9, 1., 1.1\}$. The results from the grid search algorithm are presented in table 2.

## 4. Results discussion

As can be clearly seen, in both models parameters experience a notable change after report release. Many tickers halved or doubled the trend component, and some even changed the direction of the drift. Additionally, some of the tickers experienced a notable change in volatility after the report release. The jump component appeared to be relevant in several cases meaning that the prices indeed experienced a sharp immediate change after reports' publication. Since the elasticity of the variance parameter happens to change in the breakpoint, the price sensitivity is affected as well.

To conclude, we have empirically verified the effect of quarterly reports on the stock price and acknowledged the presence of a disruptive event in the series. Therefore, time-series models are no longer applicable for forecasting, and classical statistical learning models with cross-sectional data should be used instead.